\documentclass[12pt]{article} 

\usepackage{graphicx}
\usepackage{amsmath}
\usepackage{citesort}
\usepackage{psfig}
\usepackage{psfrag}

\def\bea{\begin{eqnarray}}
\def\eea{\end{eqnarray}}
\def\ba{\begin{array}}
\def\ea{\end{array}}

\def\@authoraddress{}

\begin{document}


\title{Self-similar space-filling packings in three dimensions}

\author{Reza Mahmoodi Baram, Hans J. Herrmann}

\maketitle
\vspace{-1cm}
{\centering \small \it Institute for Computational Physics, University of        
Stuttgart,\\
Pfaffenwaldring 27, 70569 Stuttgart, Germany\\
reza@ica1.uni-stuttgart.de,
hans@ica1.uni-stuttgart.de\\}

\begin{abstract}
We develop an algorithm to construct new self-similar space-filling packings of
spheres. Each topologically different configuration is characterized by its
own fractal dimension. We also find the first bi-cromatic packing known up to
now.
\end{abstract}


\section{INTRODUCTION}
Apart from their geometrical beauty, self similar
space filling packings of spheres are used as models for ideally dense
granular packings. For
example, although we cannot expect to reproduce experimentally 
the precise positions of their spheres, 
the size distribution of such packings are good candidates to be
used in making high performance concrete, if 
a grain of sand can be approximated as a spherical particle.	   

\begin{figure} 
\begin{center}
\includegraphics[width=0.5\textwidth]{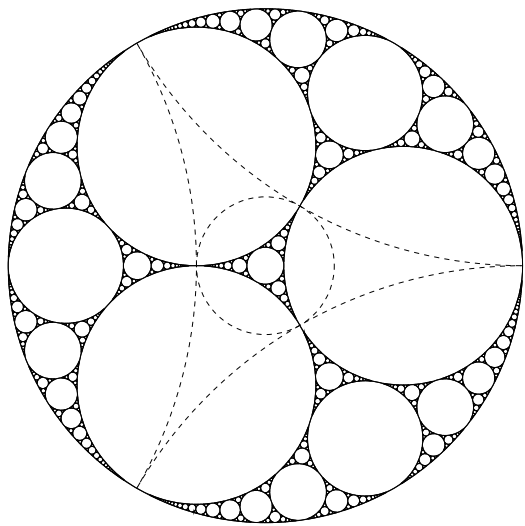}
\label{apollonian}
\caption{Apollonian packing of circles. The dashed circles are the inversion
circles.}
\end{center}
\end{figure}

The well-known Apollonian packing of circles is a two-dimensional example of
such packings (see Fig.\ref{apollonian}). In two dimensions, Herrmann et al
\cite{hans-prl,hans-gen}
have developed an algorithm to produce a variety of different packings of
circles among which the simple Apollonian packing is a special case.

Only one space filling packing in three dimension
had been constructed and studied before (see for example
Ref.\cite{Boyd1973-1,Peikert1994}). Peikert et
al\cite{Peikert1994} use a quite efficient method called {\em inversion algorithm} 
to produce this three dimensional Apollonian packing. The inversion algorithm 
is based on a simple conformal  
transformation, namely, inversion with respect to
spheres\cite{Mandelbrot}.

In this work, we show in detail how the inversion algorithm can be 
 adapted to make four more topologically different 
packings of spheres, including a packing with the important property of having 
only two classes of spheres such that no spheres from the same class touch each
other. We refer to this packing as the {\em bichromatic} packing which will
be studied in more detail in a separate paper\cite{rolling}. 
 We also calculate the fractal dimensions of new packings. Different fractal
dimensions imply the topological difference of structures of the packings. 

In Sec.\ref{sec2} we explain in detail the inversion algorithm and two ways of its generalization in producing packings
of circles. In Sec.\ref{sec3} we go over to three dimensions and discuss all 
possible packings of spheres which can be obtained using our method. Finally, in Sec.\ref{sec4} we calculate the fractal
dimensions of the obtained packings and of two dimensional cuts of the bichromatic
packings. 

\section{PACKINGS in TWO DIMENSIONS}\label{sec2}
Figure\ref{apollonian} illustrates how the inversion algorithm can be employed to
construct the classic Apollonian packing of circles within an enveloping circle of unity
radius. Initially three mutually touching circles are inscribed
inside a circular space which is to be filled. Four inversion circles are
set such that each of them is perpendicular to three of the four circles
(three initial circles and the enveloping unit circle.) 
\begin{figure} 
\begin{center}
\psfrag{00}[c][c][.7]{$C_0$}
\psfrag{00p}[c][c][.7]{$C_0'$}
\psfrag{00pp}[c][c][.7]{$C_0''$}
\psfrag{11}[c][c][.7]{$C_1$}
\psfrag{22}[c][c][.7]{$C_2$}
\psfrag{33}[c][c][.7]{$C_3$}
\psfrag{44}[c][c][.7]{$C_4$}
\psfrag{55}[c][c][.7]{$C_5$}
\psfrag{66}[c][c][.7]{$C_6$}
\psfrag{11p}[c][c][.7]{$C_1'$}
\psfrag{22p}[c][c][.7]{$C_2'$}
\psfrag{33p}[c][c][.7]{$C_3'$}
\psfrag{44p}[c][c][.7]{$C_4'$}
\psfrag{55p}[c][c][.7]{$C_5'$}
\psfrag{66p}[c][c][.7]{$C_6'$}
\includegraphics[width=0.6\textwidth]{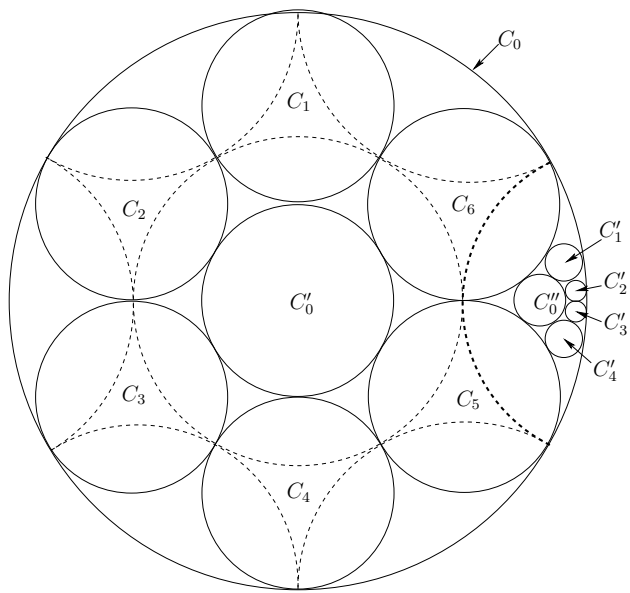}
\label{polygon}
\caption{Initial circles on the six vertices of a hexagon.}
\end{center}
\end{figure}
Beginning with this configuration, if all points {\em outside} an inversion circle 
are mapped inside, one new circle is generated, since the image of a circle
perpendicular to the inversion circle falls on itself\cite{Mandelbrot}. If the same is done for 
all the inversion circles, four new circles are generated inside the 
corresponding inversion circle. We call this the first {\em generation}. For
next generations we simply continue by applying the inversions to the newly 
generated circles. As  mentioned before, the inversion is made only
from the outside to the inside of the inversion circles, otherwise the circles of previous 
generations would be generated again. In the
limit of infinite generations we obtain the well-known Apollonian
packing, in which the circular space is completely filled with circles of
many sizes. Intuitively, there are two reasons why the inversion algorithm
fills the space. On one hand, each of the inversion circles
maps a larger more filled space
into a smaller less filled space. On the other hand, every initially
unoccupied space is inside at least one inversion circle. 

The inversion algorithm in two dimensions can be generalized in two ways. 
One is to begin with initial circles on vertices of a regular polygon
of $N$ sides 
inside the circular space instead of a triangle in the case of the 
simple Apollonian packing.
Then, we will need $N+1$ inversion circles; $N$ inversion circles
perpendicular to enveloping unit circle and to 
each pair of the initial circles which share the same side
of the polygonal and one inversion circle in the center 
perpendicular to all initial circles (see Fig.\ref{polygon}.) 
This generalization will inspire us
also in generalizing the three dimensional inversion algorithm.

\begin{figure}

\begin{center}
\includegraphics[width=0.45\textwidth]{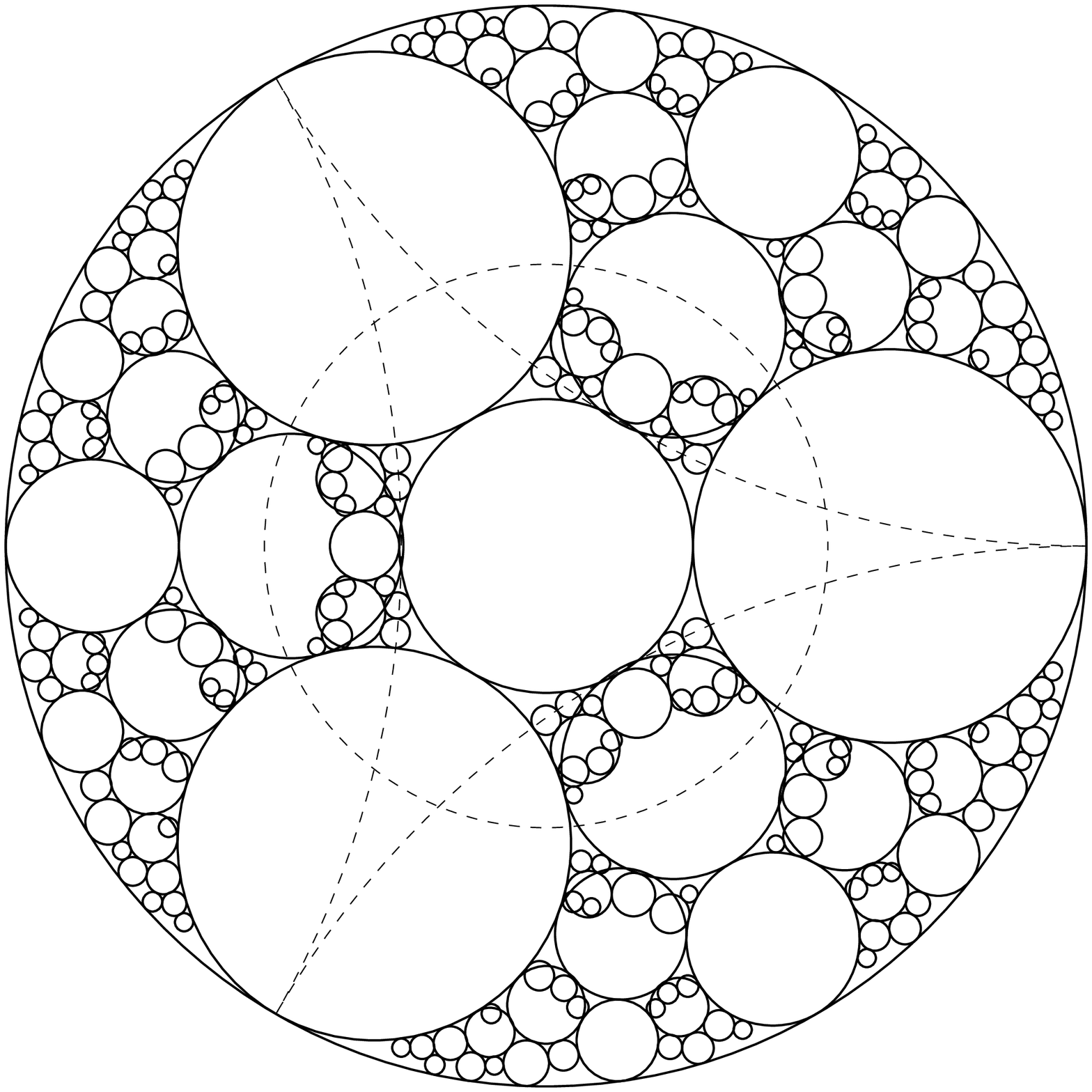}
\includegraphics[width=0.45\textwidth]{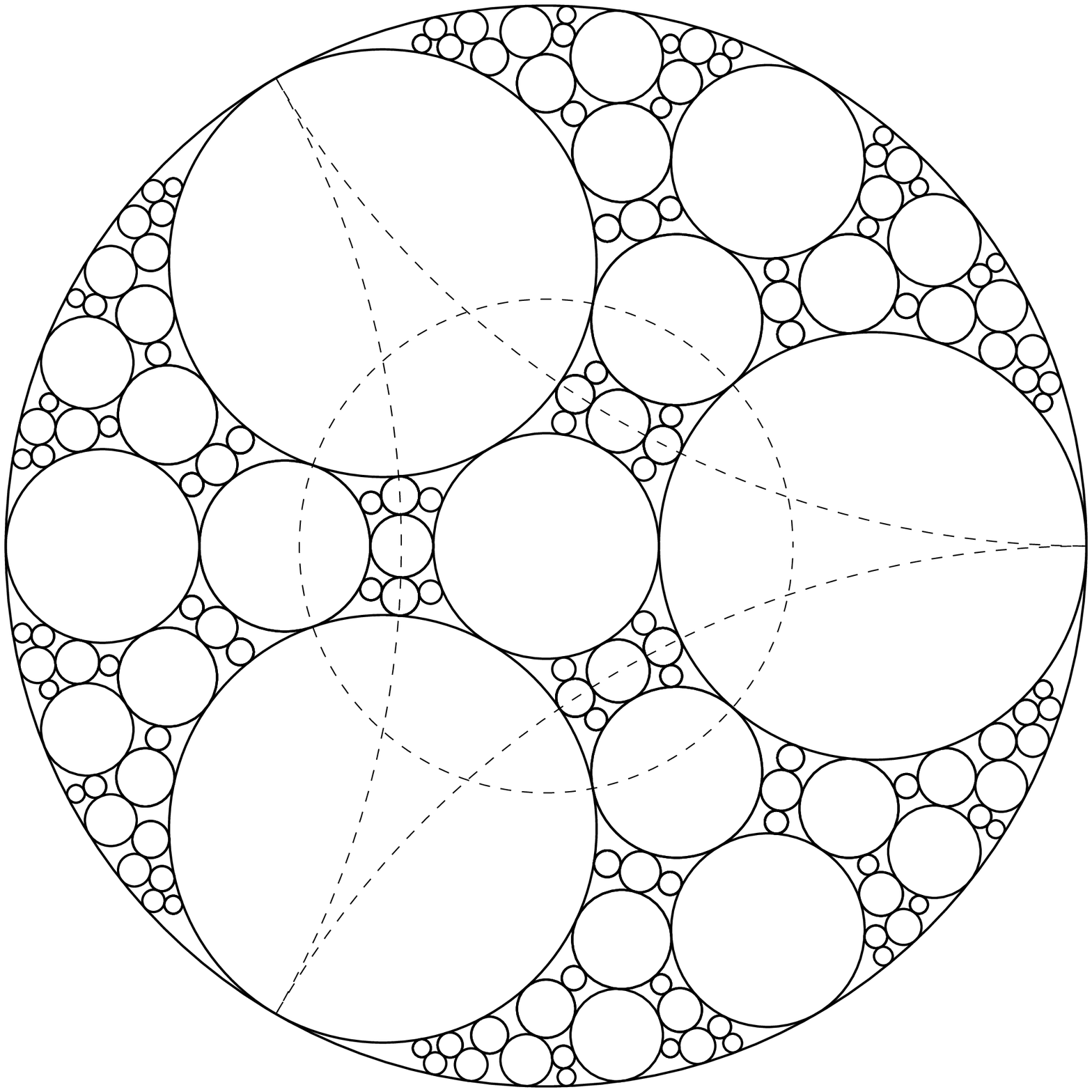}
\caption{The generalization of the simple Apollonian packing to non-touching
initial circles. If the angle between the inner and the outer inversion
circles is chosen according to Eq.(\ref{alphas}) a packing without
overlapping is obtained
(right
image, corresponding to $l=0$), otherwise no physical packing is obtained (left
image.) }\label{genAppo}
\end{center}
\end{figure}

The second way is to begin
with {\em non-touching} initial circles. We do this by reducing 
their sizes\footnote{They will be still touching 
the border of the circular space, since otherwise the condition of having 
no initially unoccupied space outside all the inversion circles is not fulfilled}. The outer
inversion circles are same as before. But, the inner 
inversion circle should be larger in order to still be perpendicular to 
the initial
circles. This, however, causes overlapping of inner inversion circles
with outer ones. This might cause a problem, because 
the overlapping regions correspond to two different
larger regions which are not necessarily mapped exactly on top of each
other. Fig.\ref{genAppo} on the right shows the case when the size of the initial circles
is chosen properly (as is explained in following) and on the left the case when size of the initial is
arbitrarily chosen. As can be seen, in the later case no physical packing is
obtained.

\begin{figure}[tb]

\psfrag{C_0}[c][c][.7]{$C_0$}
\psfrag{AA}[c][c][1.0]{$\alpha$}
\begin{center}
\includegraphics[width=0.5\textwidth]{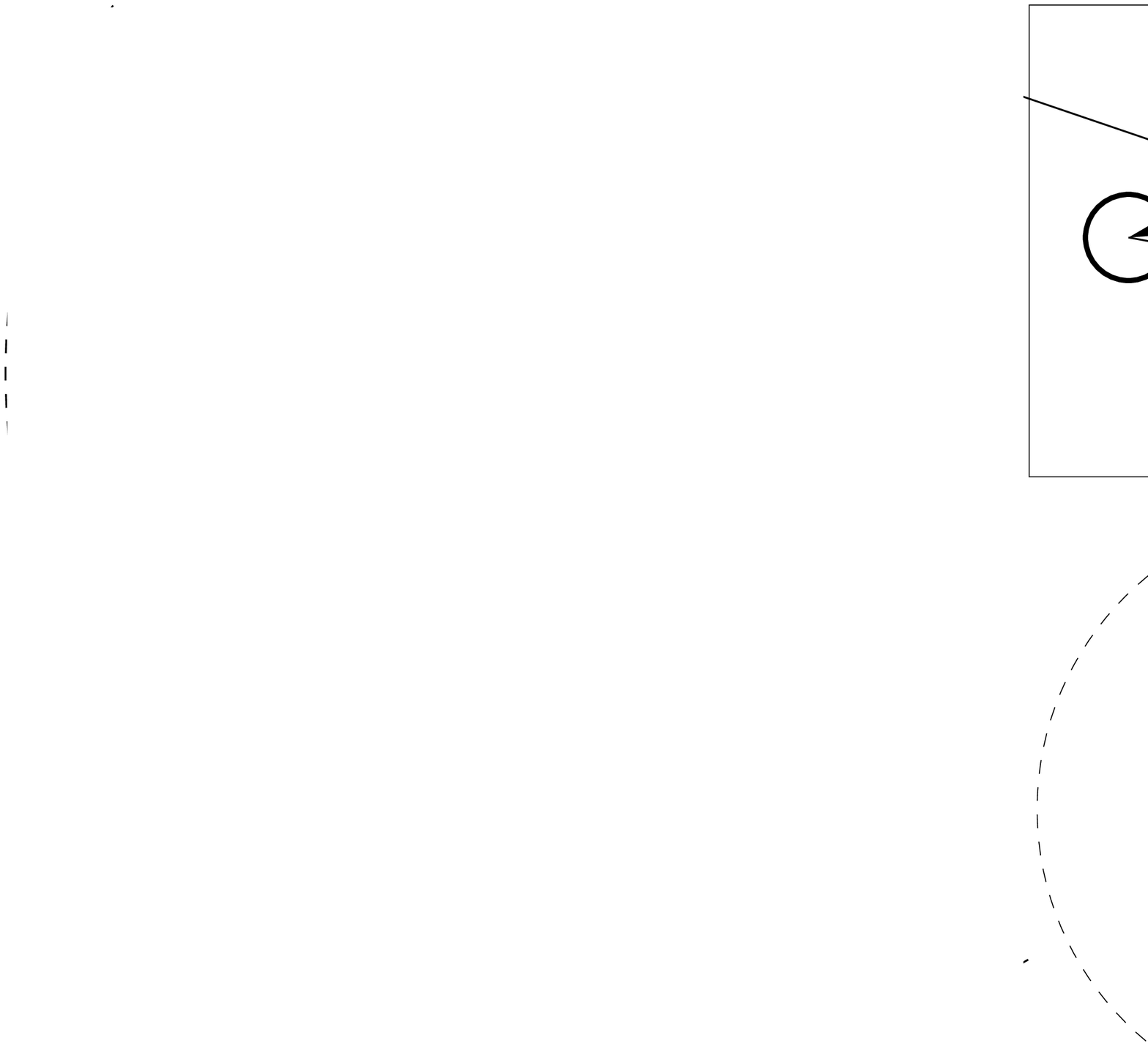}
\includegraphics[width=0.4\textwidth]{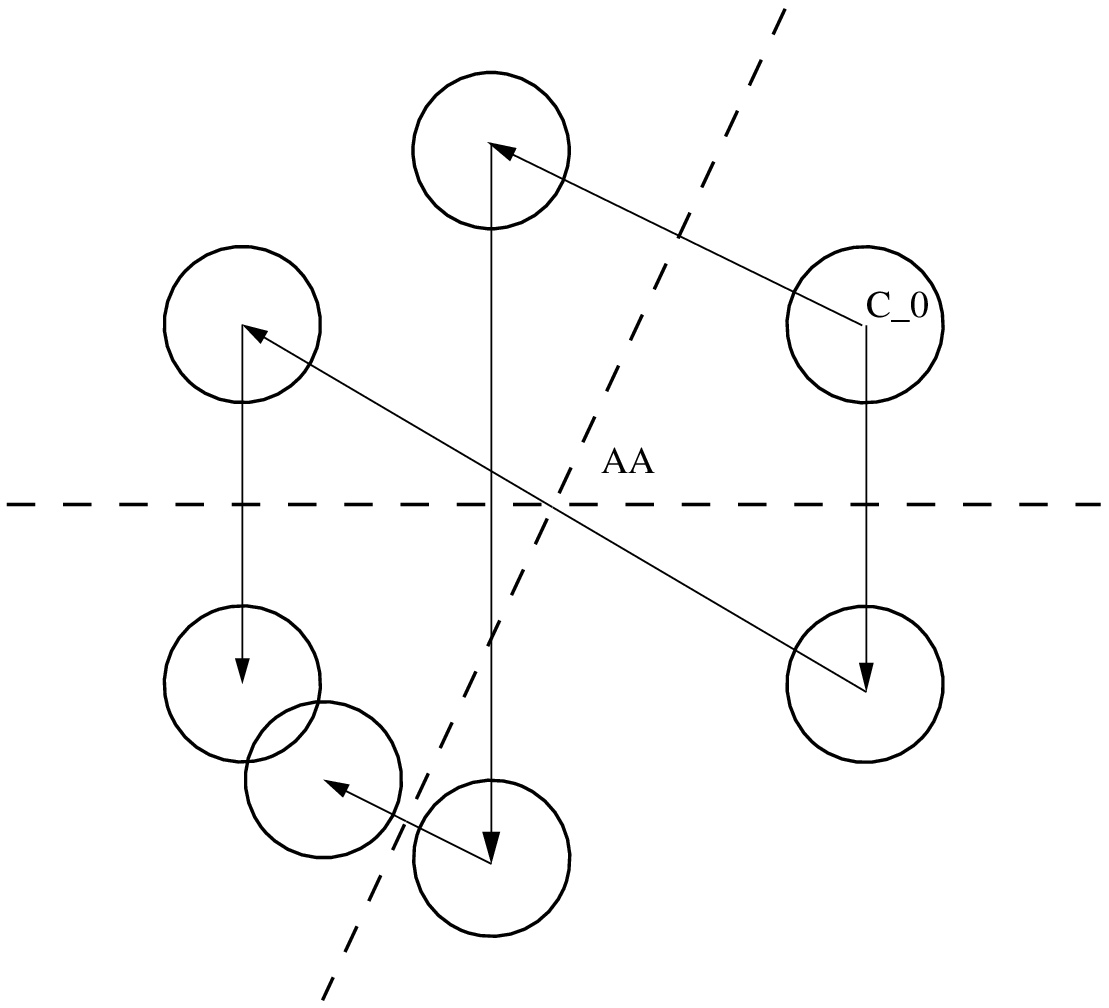}
\vspace{1cm}
\caption{Iterative inversions of a circle $C_0$  with respect to two overlapping
inversion circles (dashed lines). Two figures are equivalent (see the text).}
\label{overlapping}
\end{center}
\end{figure}

To handle this problem, we take a closer look at what happens in a sequence
of alternating inversions of an arbitrary circle $C_0$ with respect to 
two overlapping inversion circles. 
This is shown in Fig.\ref{overlapping} on the left. Beginning with one of the inversion circles, 
we obtain an image for the circle $C_0$ inside the overlapping region, 
which is not necessarily the same as the one which is obtained beginning with the other
inversion circle. However,
for some certain values of $\alpha$, which is the angle between the tangents of
the inversion circles in their intersection point, two images fall exactly on top of
each other. To find these values, we transform the whole configuration shown on the
left of Fig.\ref{overlapping} by an inversion with respect to a
circle centered around one of the intersecting points of two inversion circles.
The result is shown on the right of the figure. The inversion circles
become straight lines and act like mirrors. One can see that
under the condition  

\bea
\alpha=\frac{\pi}{l+3}, ~~~~ l=0,1,2,3\cdots \label{alphas}
\eea
two images fall exactly on top of each other\footnote{ $\alpha=\frac{\pi}{2}$
corresponds to the case when the size of initial circles is zero and the
circular space is completely filled by {\em only one}
circle.}.

Considering Eq.(\ref{alphas}) in choosing the size of initial
circles, we obtain an infinite number of different packings corresponding to the
different values of $l\geq 0$ and $N\geq 3$. The only remaining problem is that each
generated circle in the overlapping
region is generated more than once, which can be easily solved by eliminating 
the repeated images. The packing with $l=0$ and  $N=3$ is shown in Fig.\ref{genAppo} on the
right, together
with the case which doesn't fulfill Eq.(\ref{alphas}) on the
left. The classic Apollonian packing of circles shown in the
Fig.\ref{apollonian} corresponds to $l=\infty$ and $N=3$.

The packings we obtain here turn out to be the same as those constructed  
and named as first family of space-filling bearings by 
Herrmann et al \cite{hans-prl} with $m=N-3$ and $n=l$ in their nomenclature. 
Fig.\ref{genAppo} (on the right) is corresponding to $m=0$, $n=0$ as they classify it.
All these packings are deterministic, self-similar fractal with a dimension 
which is different for different $n$ and $m$.

One may note that Eq.(\ref{alphas}) should also hold in three dimensions when
we deal with overlapping inversion spheres, since the cross section of any
plane passing through the centers of the spheres gives the same situation 
as in two dimensions.

\section{GENERALIZATION to THREE DIMENSIONS}\label{sec3}
As was mentioned before, Peikert et al  used the inversion algorithm 
to very efficiently make the, so-called, three dimensional 
Apollonian packing of spheres within an enveloping sphere of radius one. In
this section, we use this algorithm and its generalization discussed in the
previous section to produce more such three dimensional self-similar space-filling
packings in three dimensions.

Peikert et al  begin with four initial spheres on the vertices of a tetrahedron inside a
unit sphere which is to be filled. They placed five inversion spheres, four
spheres perpendicular to the enveloping unit sphere and the three initial spheres
belonging to each face of the tetrahedron, and one in the center
perpendicular to all four initial spheres. However, the inversion spheres, even in this
case which is the simplest three dimensional configuration, do overlap. We
come back to this later in this section. 

The fact that in two dimensions we can replace a triangle by other regular polygons
as the basis for new configurations of initial circles suggests to do a similar
extension in three dimensions. So, we examine the configurations based on
the other {\em Platonic Solids} besides the tetrahedron as possible candidates. 
In this section, we will study all these cases and their resulting
packings. An important restriction is, however, that the number of 
Platonic Solids is limited
to five. Furthermore, as we will see in the following, the 
inversion spheres overlap in all
configurations. Therefore we do not expect as many packings
as in two dimensions. Actually, we will end up with only
five packings. 

\begin{figure}[tb]
\begin{center}
\includegraphics[width=1\textwidth]{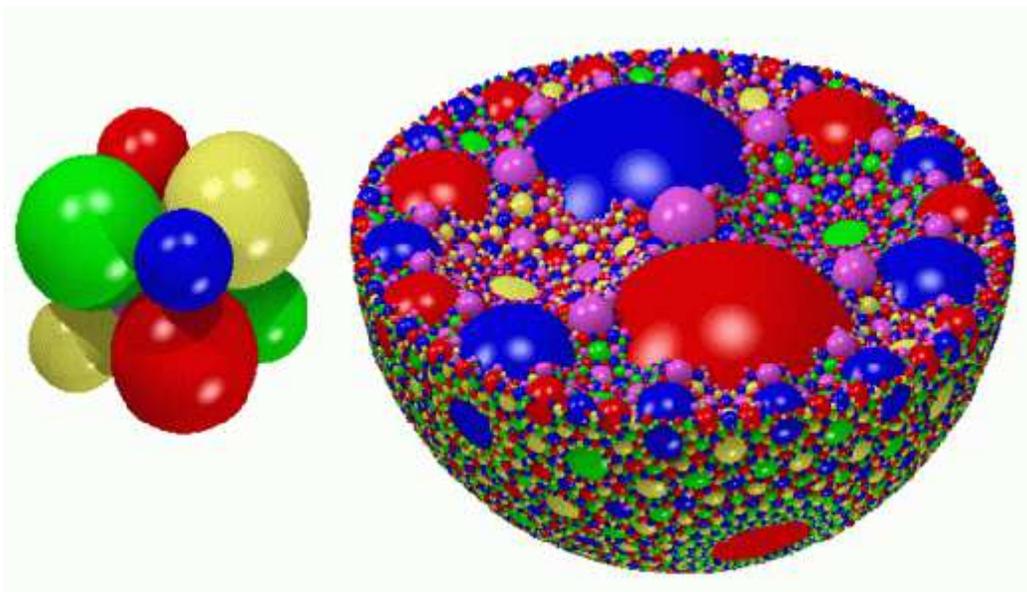}
\caption{The tetrahedron-based packing.}
\label{tetra}
\end{center}
\end{figure}

We begin with mutually touching initial 
spheres on the vertices of a Platonic Solid 
inside the unit sphere. 
Inversion spheres are placed as follows; 
One in the center of the unit
sphere perpendicular to all initial spheres, and one at each face 
of the Platonic Solid perpendicular to the enveloping unit sphere and the 
initial spheres which form that face of the Platonic Solid. 
Using this configuration of initial and inversion spheres, the process 
of filling the space is exactly the same as explained in the last section, that is, 
iteratively mapping the initial spheres into smaller and smaller unoccupied 
spaces. Figure \ref{tetra} shows the packing based on the tetrahedron. The image on
the left shows the packing after {\em one} generation and the one on the
right shows the same packing including all spheres with radii larger
than $2^{-7}$. The spheres are grouped into different classes (assigned
by different colors) such that no spheres having the same color touch each
other. For the tetrahedron-based packing there are five such 
classes of spheres, as can be seen in the figure.

\begin{figure} [tb]
\psfrag{AA}[c][c][.7]{$\alpha$}
\begin{center}
\includegraphics[width=0.5\textwidth]{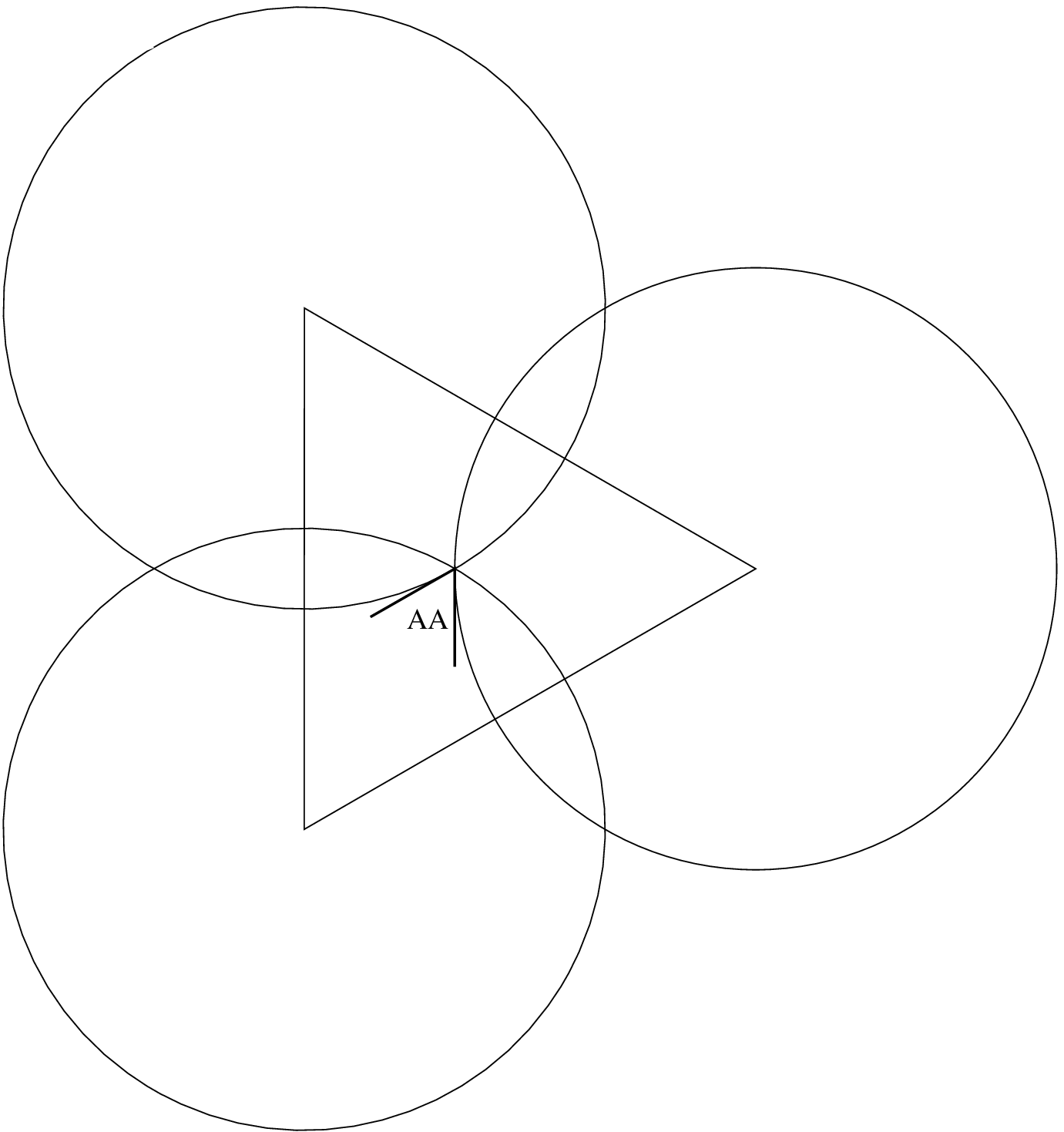}
\caption{Overlapping of the inversion spheres in the case of tetrahedron.
$\alpha$ equals$\frac{\pi}{3}$.}
\label{inv-overlapping-tetra}
\end{center}
\end{figure}

As we discussed in the last section, in order that the iterative inversions 
lead to an allowed packing, the inversion spheres 
should either not overlap or, if they do, the angle $\alpha$ 
between them should follow Eq.(\ref{alphas}). In this case $\alpha$ is the
angle between the planes tangent to two inversion spheres at their
intersetion points. 
We check the value of $\alpha$ for each Platonic Solid. To do this, we note that the 
centers of the inversion
spheres are on the vertices of the {\em conjugate Platonic Solid}.
As an example, Fig.\ref{inv-overlapping-tetra} shows the plane containing 
one of the faces of the Platonic Solid conjugate to the tetrahedron 
(which is also a tetrahedron). This plane is
tangent to the unit sphere and to one of the initial spheres at their
contact point. The circles are the
cross section of the plane with the inversion spheres. 
One can easily see that the angle between two neigbouring inversion spheres is
$\frac{\pi}{3}$, which happens to be equal to the angle corresponding to 
$l=0$ in Eq.(\ref{alphas}). 
The corresponding angle between the inner and the outer 
inversion spheres can be also 
calculated and is $\frac{\pi}{3}$ too. 
Therefore an allowed configuration will be constructed. Because with this
angle the overlapping of the invertion spheres will cause no problem but only
multi-generation of spheres in overlapping region. As in
last section, this problem is solved by eliminating the repeated spheres. 
\begin{figure}[b]
\begin{center}
\includegraphics[width=1\textwidth]{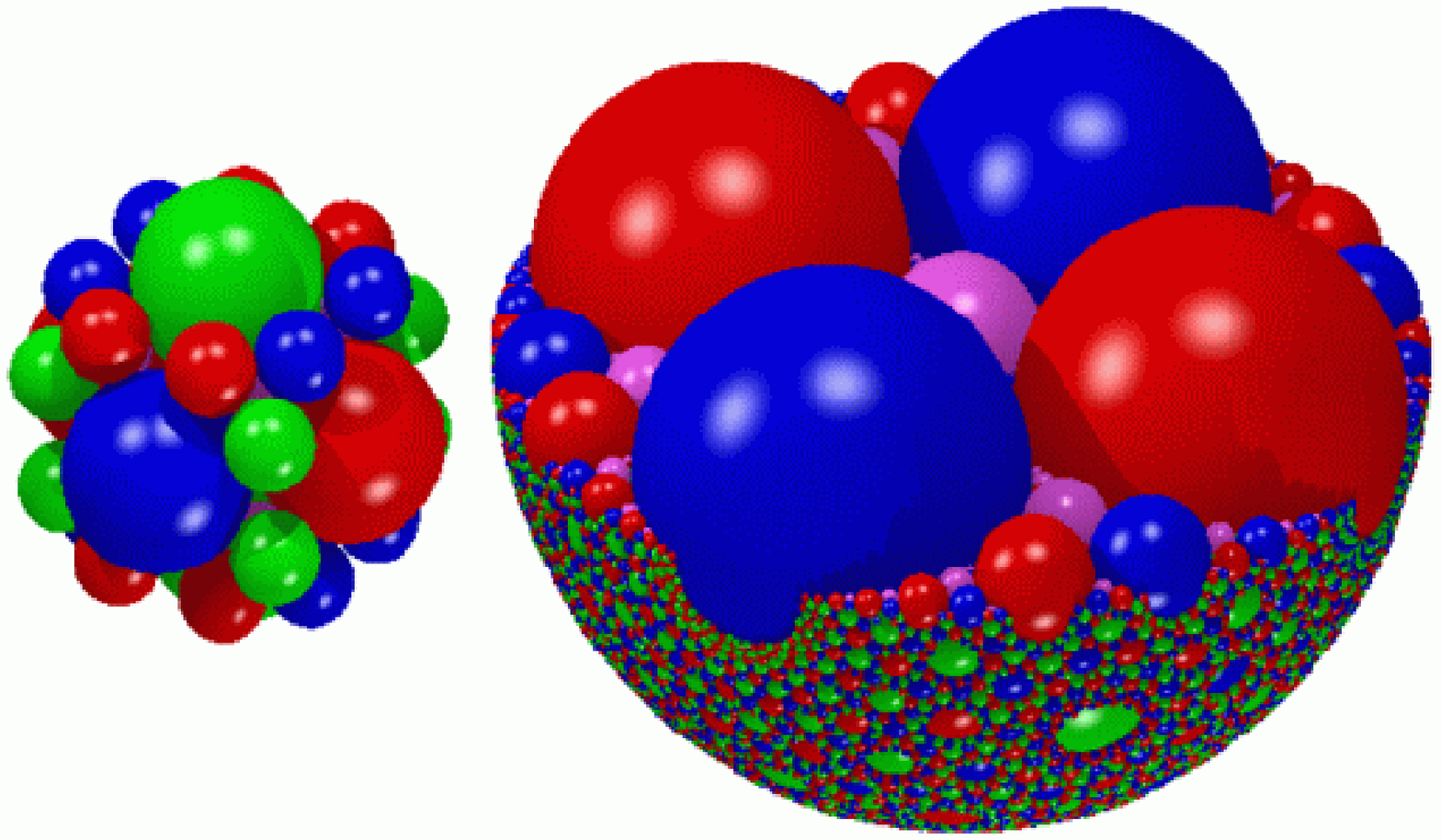}
\caption{The octahedron-based packing.}
\label{octa1}
\end{center}
\end{figure}
\begin{figure}[b]
\begin{center}
\includegraphics[width=1\textwidth]{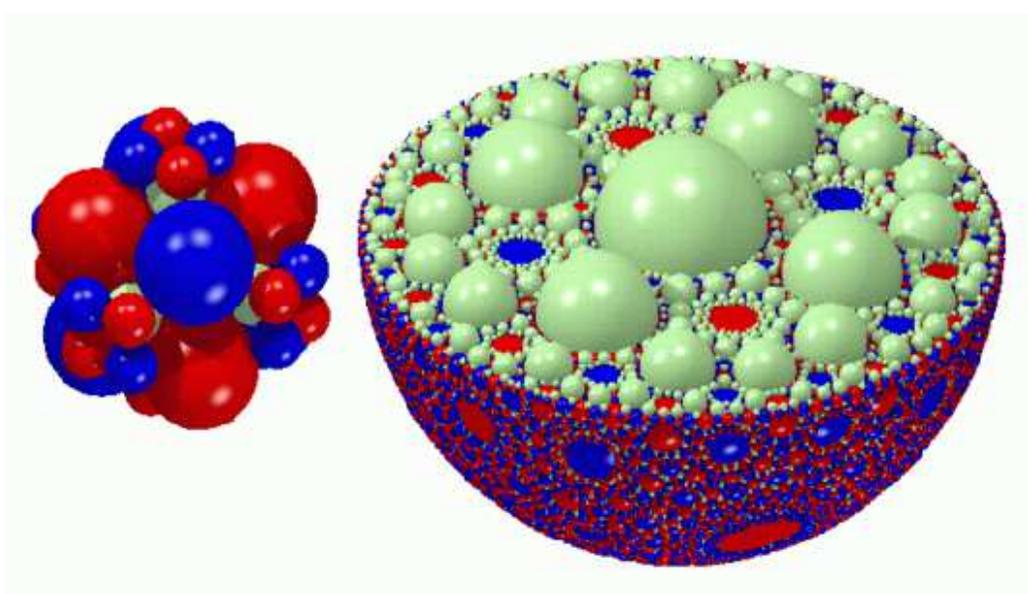}
\caption{The cube-based packing.}
\label{cube}
\end{center}
\end{figure}
\begin{figure}[b]
\begin{center}
\includegraphics[width=1\textwidth]{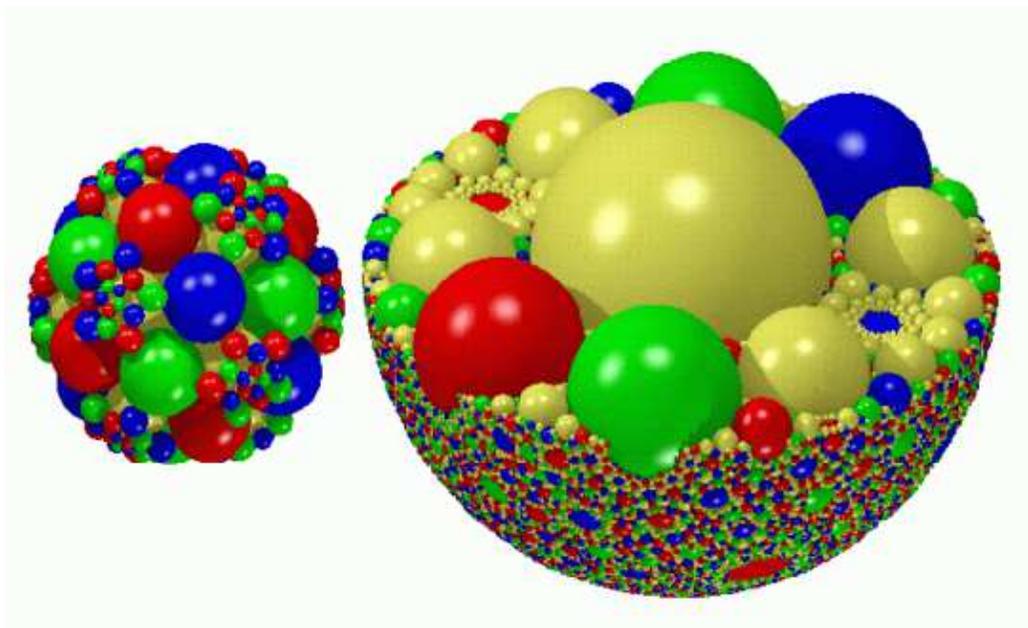}
\caption{The dodecahedron-based packing.}
\label{dodeca}
\end{center}
\end{figure}
\begin{figure}[b]
\begin{center}
\includegraphics[width=1\textwidth]{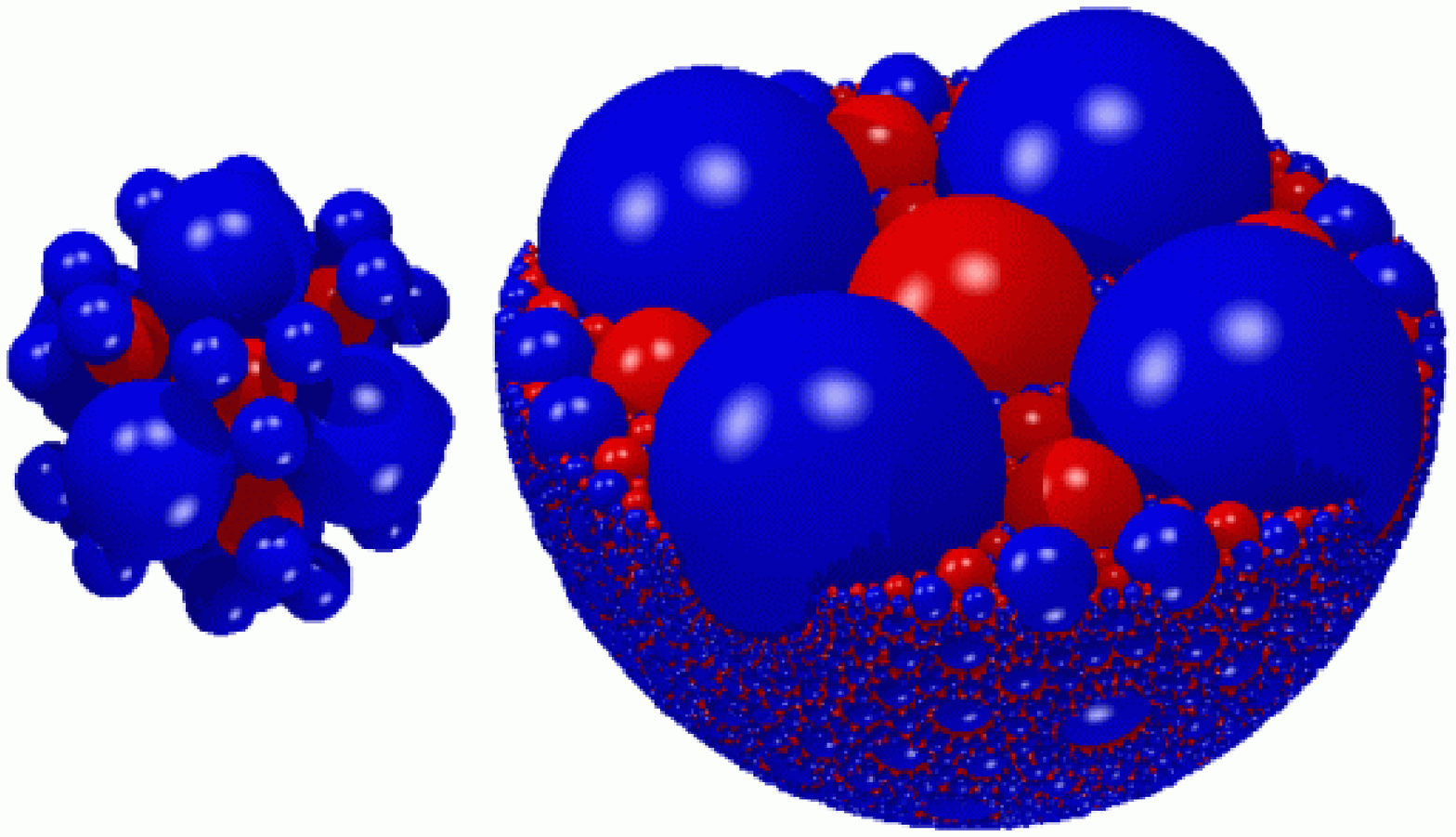}
\caption{The bichromatic packing which is the second octahedron-based packing.
No spheres of the same colour touch each other.}
\label{bichromatic}
\end{center}
\end{figure}

The first new packing is based on the octahedron, and in this case the angles
between ovelapping inversion spheres again follow 
Eq.(\ref{alphas}), having the value $\frac{\pi}{4}$ (note that the
conjugate of an octahedron is a cube whose faces are squares).  
The result is shown in Fig.\ref{octa1}. In this case, there are four
classes of spheres leading to four different colours in order that no
spheres of the same colour touch each other.  

For the case of a cube as the basis for the initial spheres, the angle between
the overlapping inversion spheres is also $\frac{\pi}{3}$ as in the case of
the tetrahedron, 
since the conjugate of a cube is an octaderon whose faces are triagular too. 
So, in this case the inversion algorithm also produces an allowed packing. 
The result is shown in Fig.\ref{cube}. In this case there are three
classes and therefore three colours of spheres.

For the same reason as the case of cube, we obtain a packing for the case of
the dodecahedron, since the conjugate of a dodecahedron is an icosahedron with
trianglar faces. The result is shown in Fig.\ref{dodeca}. In this case there 
are
four classes and therefore four colours of spheres.

\begin{figure}[bt]
\psfrag{AA}[c][c][.7]{$\alpha$}
\begin{center}
\includegraphics[width=0.5\textwidth]{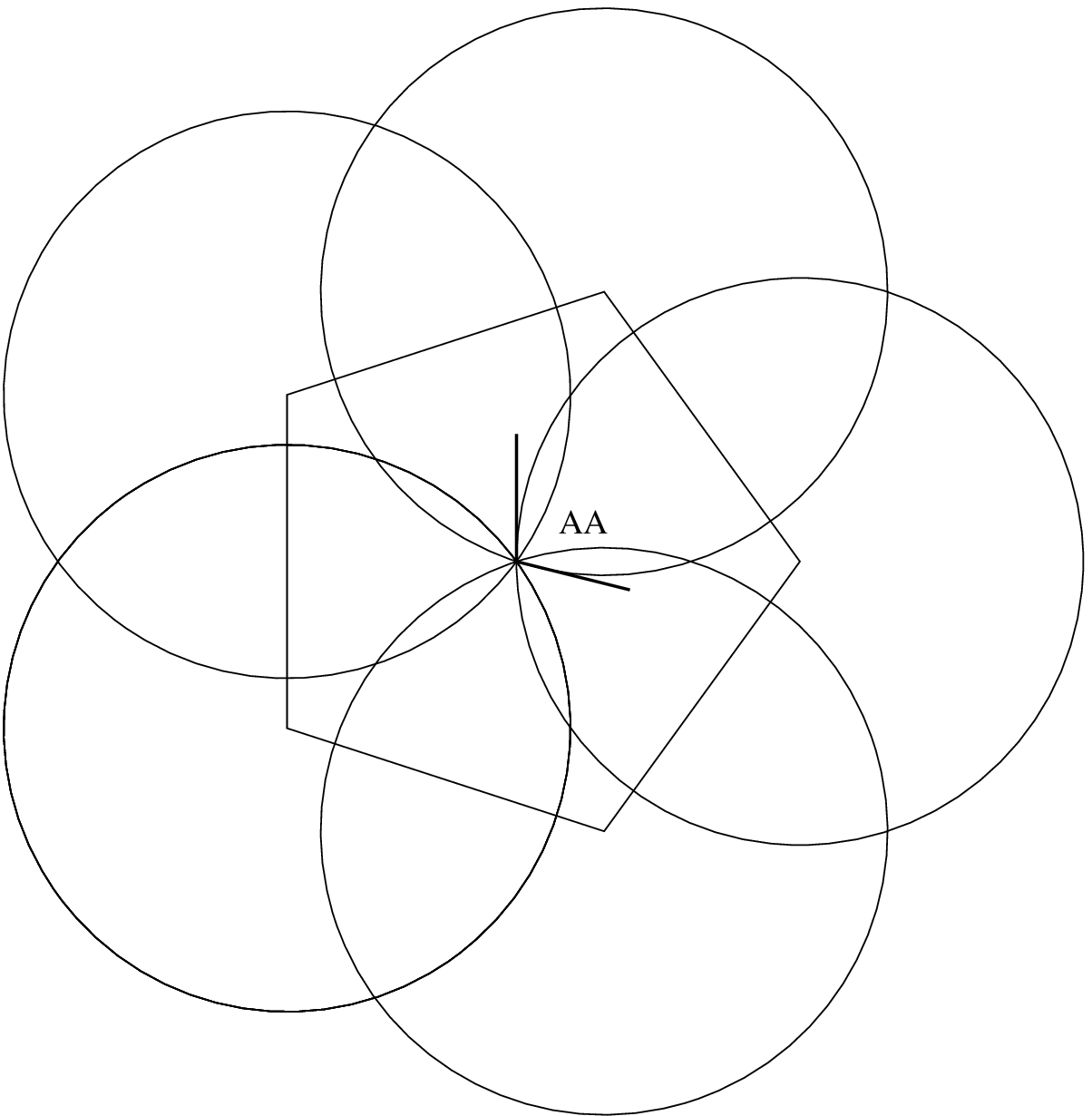}
\caption{Overlapping of the inversion spheres in the case of the icosahedron.
$\alpha$ equals $\frac{2\pi}{5}$.}
\label{inv-overlapping-icosa}

\end{center}
\end{figure}

But, for the case of icosahedron we do not find any allowed packing. 
Because as is shown
in Fig.\ref{inv-overlapping-icosa}, the conjugate of an icosahedron is a
dodecahedron with pentagonal faces. This  leads to the value $\frac{2\pi}{5}$
for the angle between neighbouring inversion spheres, which does not
match any value of $\alpha$ in Eq.(\ref{alphas}).

In the previous section we discussed the generalization of having non-touching
initial circles. This can also be
applied in three dimensions. Similar to two dimensions, reducing the size
of the initial spheres we also need to expand the inner inversion sphere
(the others are same as before) in order to
still be perpendicular to the initial spheres. But, due to the values of
$\alpha$, one can see that overlappings of the inversion spheres are at
their allowed maximum, except for the case of the octahedron for which
$\alpha=\frac{\pi}{4}$. In this case the only value of $\alpha$ which leads
to an expanded inner inversion sphere and follows Eq.(\ref{alphas}) will be
$\frac{\pi}{3}$. Therefore, this is the only
packing with non-touching initial spheres which can be constructed. 

As an important property of this packing, one can see that only two
different colours are enough  such that no spheres
of same colour touch each other, because the initial spheres don't touch each
other and only touch the sphere in the center and this geometry
is preserved in all scales, due to the self-similar nature of construction of
the packing. We call this packing bichromatic.

\begin{figure}
\begin{center}
\psfrag{Df}[c][c][0.9]{$D_f$}
\psfrag{KKK}[c][c][1.0]{$-\log \epsilon$}
\psfrag{KKK2}[c][c][0.9]{$-\log \epsilon$}
\psfrag{NNN}[c][c][1.0]{$\log N(\epsilon)$}
\includegraphics[width=0.8\textwidth]{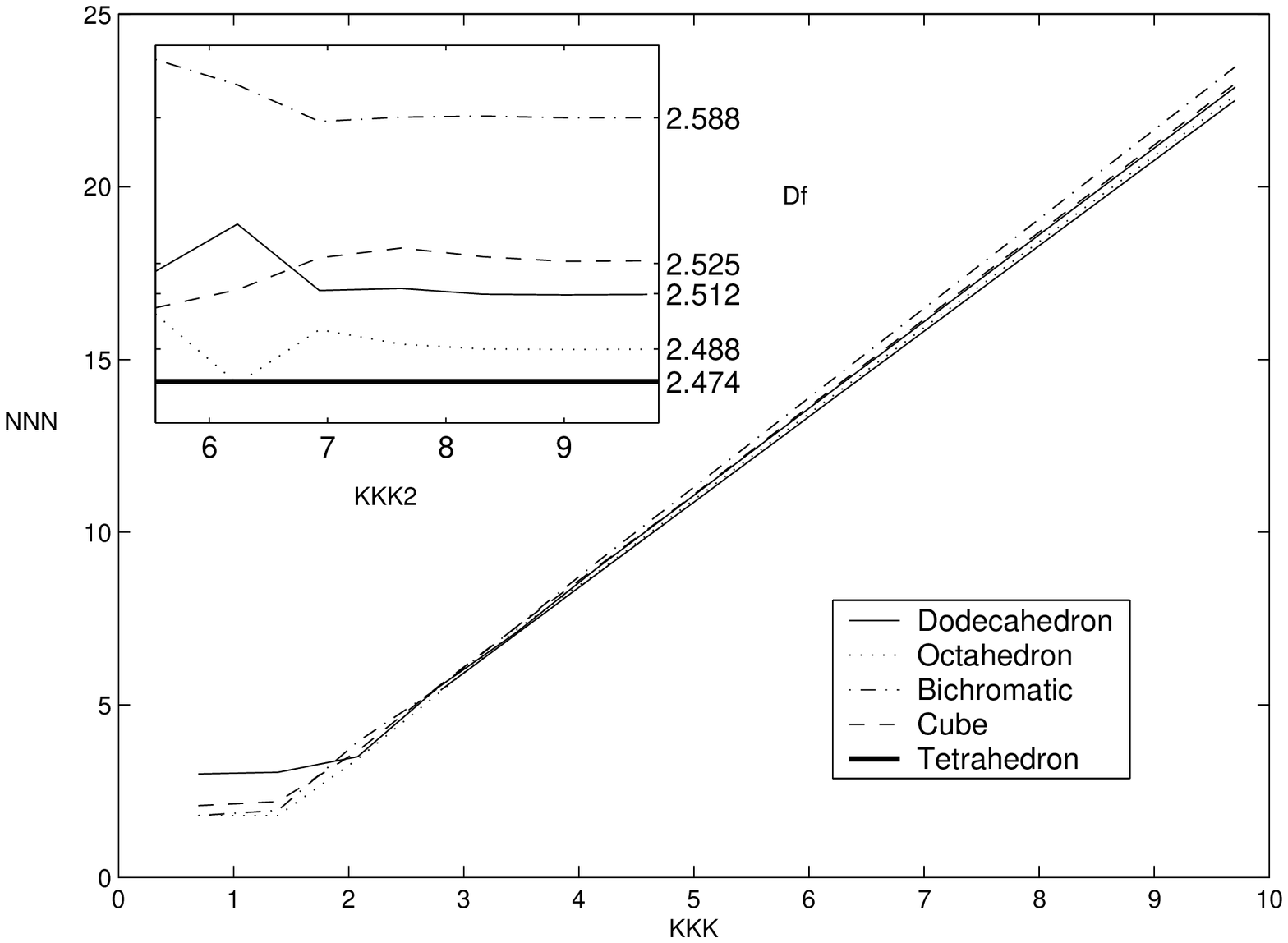}
\caption{The fractal dimension of different Packings of
spheres.}\label{platonFD}
\end{center}
\end{figure}
 
\begin{figure}
\begin{center}
\psfrag{Df}[c][c][0.9]{$D_f$}
\psfrag{KKK}[c][c][1.0]{$-\log \epsilon$}
\psfrag{KKK2}[c][c][0.9]{$-\log \epsilon$}
\psfrag{NNN}[c][c][1.0]{$\log N(\epsilon)$}
\includegraphics[width=0.8\textwidth]{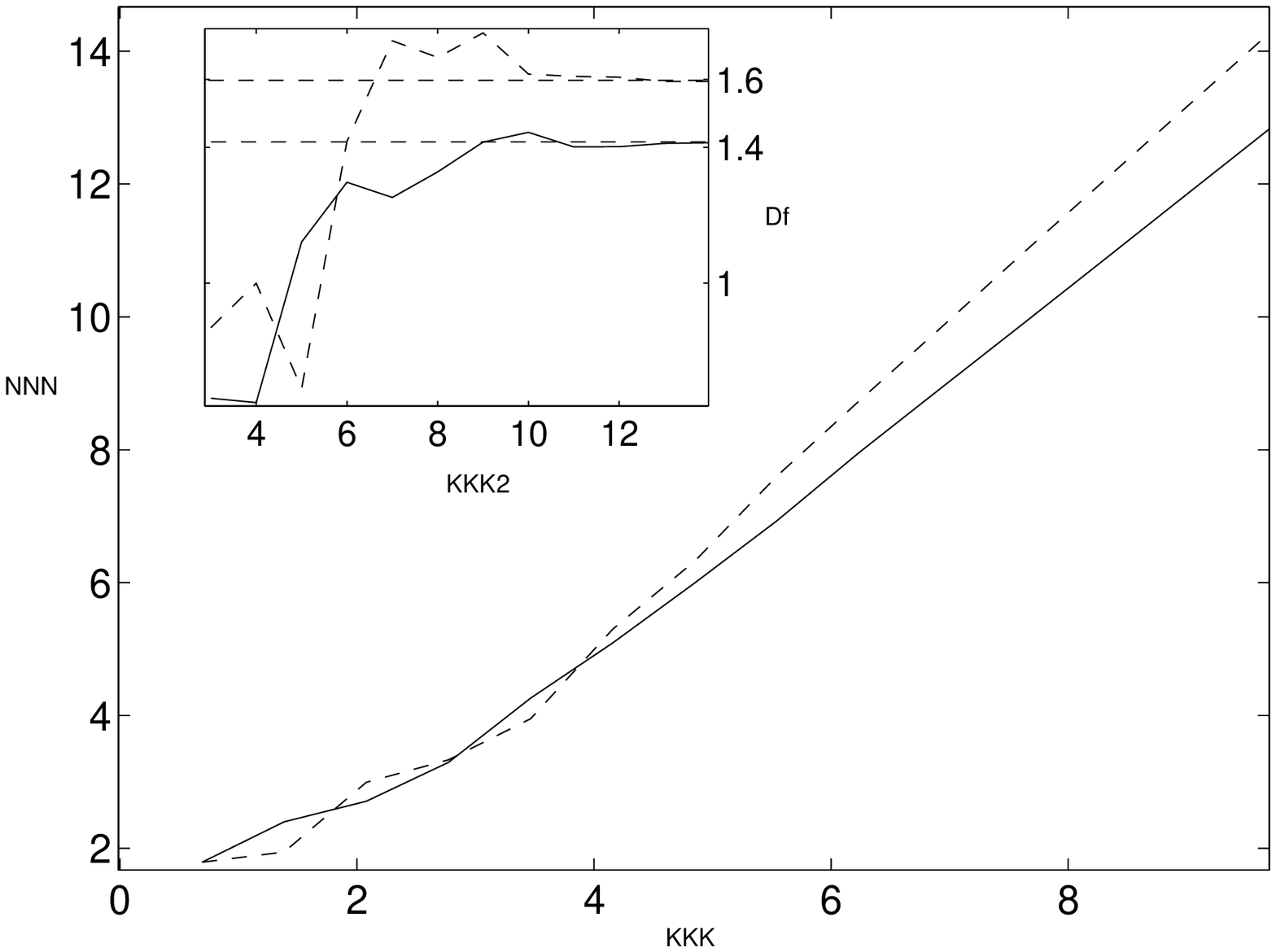}
\caption{The fractal dimension of two cuts of the bichromatic
packing.}\label{cutFD}
\end{center}
\end{figure}

\section{FRACTAL DIMENSIONS}\label{sec4}

As we mentioned before, the obtained packings have self-similar structures,
and, therefore, are fractals. The method we use to calculate the 
fractal dimensions is the same as the one
used in ref.\cite{Peikert1994}. The number of all spheres in a packing with 
radii larger than $\epsilon$ follows an asymptotic relation\cite{Boyd1973-2,Boyd1982}, 
\bea
N(\epsilon)\sim \epsilon^{-D},
\eea
in which $D$ is the fractal dimension of the packing.
Figure \ref{platonFD} shows $N(\epsilon)$ in the logarithmic
scales for different packings. One can see that, for smaller $\epsilon$'s
the curves become linear whose slope determines the fractal dimension of the
corresponding packing. As shown in the inset of the figure, the slope is
calculated by linear fitting to an interval containing five
points and shifting this interval towards smaller $\epsilon$'s. 
The difference between the fractal dimensions 
reflects the topological difference of the obtained packings. So, the
numerical precision is improved until this difference become evident. 
The fractal dimension of the packing based on tetrahedron has been calculated in
Ref.\cite{Peikert1994} to a high degree of precision. It is approximately
$2.474$, which is shown in the figure as the lowest value.

We also calculate the fractal dimensions of two cuts of the bichromatic
packing; one through the plane containing four vertices of the octahedron, which also passes
the center, and another parallel to this plane but $0.5$ off the center. The
result is shown in Fig.\ref{cutFD}. The difference between the fractal
dimensions of the cuts shows that the packing is not a homogeneous fractal. 

\section{CONCLUTION}
We developed an algorithm for three dimensional self-similar space-filling packings of
spheres and constructed four till-now-unknown realizations. 
These packings are topologically  different
since they possess different fractal dimensoins. Among them there
is a packing which can be represented with only two colours such that no spheres
of the same colour touch each other. This is a significant property since it
leads to the possibility of rotating spheres with no frustration and
therefore to space-filling bearings of spheres, as will be shown in a
separate paper (ref.\cite{rolling}).

\section{ACKNOWLEDGMENT}
We would like to thank R. Peikert for explaining their algorithm to us and
giving us the code.
 
\bibliographystyle{unsrt}

\end{document}